\documentclass[conference]{IEEEtran}
\IEEEoverridecommandlockouts
\usepackage{cite}
\usepackage{amsmath,amssymb,amsfonts}
\usepackage{algorithmic}
\usepackage{graphicx}
\usepackage{textcomp}
\usepackage{xcolor}
\def\BibTeX{{\rm B\kern-.05em{\sc i\kern-.025em b}\kern-.08em
    T\kern-.1667em\lower.7ex\hbox{E}\kern-.125emX}}
\begin{document}

\title{UVM Based Reusable Verification IP for Wishbone Compliant SPI Master Core}

\author{\IEEEauthorblockN{Lakhan Shiva Kamireddy\IEEEauthorrefmark{1}, Lakhan Saiteja K\IEEEauthorrefmark{2}}
\IEEEauthorblockA{\IEEEauthorrefmark{1}VLSI CAD Research Group, Department of Electrical and Computer Engineering, University of Colorado\\
Boulder, CO 80303, USA, Email: lakhan.kamireddy@colorado.edu}
\IEEEauthorblockA{\IEEEauthorrefmark{2}Department of Electronics and Electrical Communication Engineering,
Indian Institute of Technology Kharagpur\\ West Bengal 721302, India, Email: lakhansaiteja@gmail.com}
}

\maketitle

\begin{abstract}
The System on Chip design industry relies heavily on functional verification to ensure that the designs are bug-free. As design engineers are coming up with increasingly dense chips with much functionality, the functional verification field has advanced to provide modern verification techniques. In this paper, we present verification of a wishbone compliant Serial Peripheral Interface (SPI) Master core using a System Verilog based standard verification methodology, the Universal Verification Methodology (UVM). By making use of UVM factory pattern with parameterized classes, we have developed a robust and reusable verification IP. SPI is a full duplex communication protocol used to interface components most likely in embedded systems. We have verified an SPI Master IP core design that is wishbone compliant and compatible with SPI protocol and bus and furnished the results of our verification. We have used QuestaSim for simulation and analysis of waveforms, Integrated Metrics Center, Cadence for coverage analysis. We also propose interesting future directions for this work in developing reliable systems.
\end{abstract}


\section{INTRODUCTION}


With the ever-increasing complexity of designs, system level verification of large System on Chips (SoC's) has become complex [1]. These design complexities are accompanied with the interdependencies of various functionalities, which make the design more susceptible to bugs. Efficiently verifying the designs and reducing the time to market without compromising on the targets of achieving bug-free designs is a herculean challenge to verification teams. Functional verification is a process of ensuring that a design performs the tasks as intended by the overall system architecture. With monotonically increasing costs of re-spins and requiring additional manpower as well as development time, verification is the most critical phase in chip design cycle. It takes nearly 70\% of the total design cycle [1]. Design reuse and verification reuse are essential in today's constrained time-to-market requirements. Hence, the need to develop robust and reusable verification IP arises.

Simulation-based verification is a standard and popularly used method of functional verification. System Verilog (SV), the Hardware Description and Verification Language (HDVL) has massive enhancements over Verilog, which provides several features to develop a fully functional verification environment with support for artifacts like object oriented programming, coverage analysis, constrained randomization and assertion based VIP. A methodological approach for verification increases the efficiency and reduces the verification effort. In this paper, we use UVM, a System Verilog based methodology for testing an SPI master core that is wishbone compliant.

The paper is organized into the following sections: Section II introduces the key features of SV and UVM environment. In Section-III, we introduce the SPI Master IP core for which the UVM framework is developed. Section-IV presents our approach towards the development of UVM based VIP. Simulation results with snapshots and a critical discussion of the limitations of the design are presented in Section-V. A conclusion is drawn in Section-VI.

\section{SYSTEM VERILOG AND UVM ENVIRONMENT}
Traditional testbenches were Verilog based and were not meant to verify complex SoC designs. The verification efforts were carried out by designers themselves by merely applying a sequence of critical stimulus elements to the Device Under Test and match the result to the expected outcome. As chip size kept decreasing, chip function became more complicated, and verification effort dominated the design process. Conventional methodology proved futile in verifying complex modern designs. When Verilog was declared an IEEE standard, the Accellera Systems Initiative had come up with revised versions of Verilog [2]. To cater modern verification needs, many proprietary verification languages were developed. Consequently, System Verilog was developed by Accellera, by extending Verilog with many of the features of Superlog and C languages. Subsequently, a substantial number of verification capabilities were added to System Verilog. System Verilog also borrowed many features from the C language, C++, Vera C and VHDL with support for complete verification [2].

In [2], Peter Flake et al. presented the features of System Verilog while discussing reasons for particular language design choices. Not only with inbuilt support for object-oriented programming, but also importing features of several domain-specific languages, System Verilog stands as a popular digital HDVL. Some useful features are- code interface allowing someone reusing code to concentrate on the features the code provides, not on how the code is implemented, virtual interfaces, coverage driven constrained random verification, assertions, clocking blocks, functional coverage constructs [4].

In the evolutionary stages of System Verilog, the choice of verification methodology was strongly related to the choice of tool vendor. To achieve vendor independence, Accellera Systems Initiative took up the task of creating an open standard methodology that could be used with all major vendors' tools. The result is the well known Universal Verification Methodology. UVM is an open-source library which is portable to any HDL simulator that supports System Verilog. It provides a rich base class library thus supporting the construction and deployment of verification components (VCs) and testbenches, massively reducing the coding effort [5].

A VC is a ready to use component, which is reusable. VCs share a set of conventions and consist of a complete set of elements for stimulating, checking and collecting coverage. The stimulus provided to the DUT through a VC verifies protocol implementation and design architecture. Fig. 1 (Gayathri M 2016 [7]) shows a generic UVM testbench architecture. The top-level module in a UVM testbench comprises of a DUT, and a testbench is connected to it.

\subsection{UVM Testbench}
UVM testbench comprises of sequence item, sequencer, driver, monitor, agent, scoreboard, environment, test suite.
\begin{figure}[!t]
\centering
\includegraphics[scale=0.25]{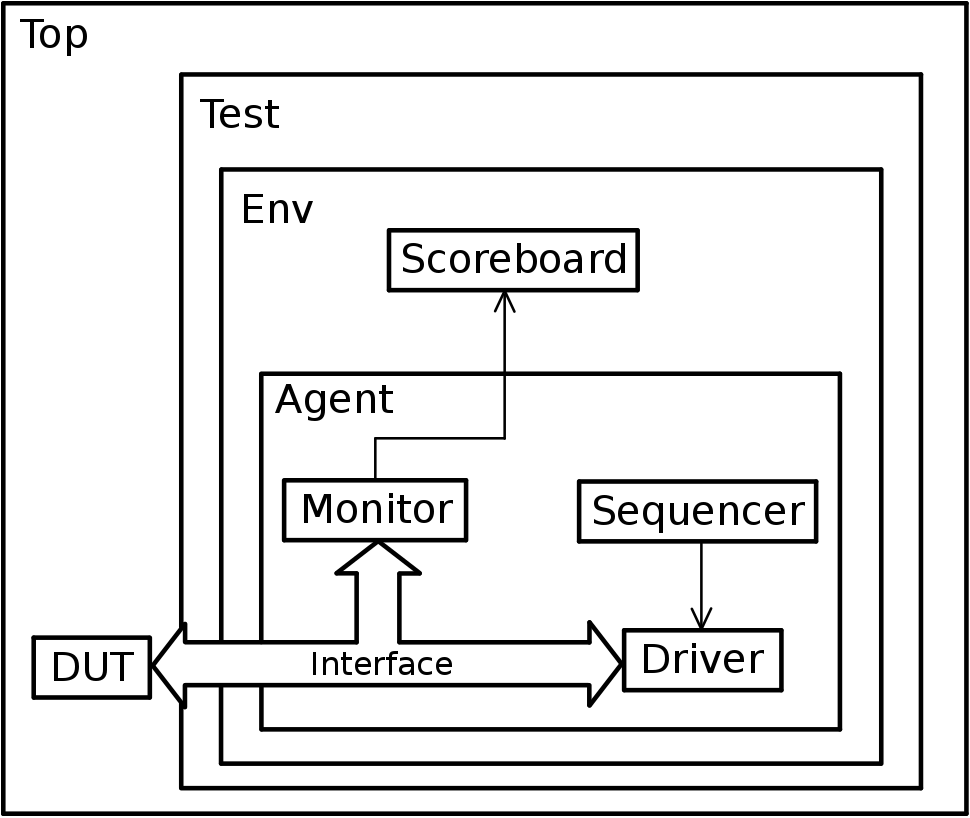}
\caption{UVM Testbench Architecture}
\end{figure}


\subsubsection{Top Testbench}
Testbench comprises of instantiations of DUT modules and interfaces that connect DUT with testbench. Transaction Level Modeling (TLM) interfaces in UVM are a great resource to implement communication function calls for transmitting and receiving transactions among modules.

\subsubsection{Test}
Test component is a class under testbench. Typical tasks performed in this are applying the stimulus to DUT by invoking sequences, configuring values in config class. Test class instantiates the top level environment.

\subsubsection{Environment}
Environment (env as illustrated above) is a reusable component class that aggregates scoreboards, agents and other UVM verification environments together.

\subsubsection{Agent}
The Agent as seen in the above illustration aggregates several verification components such as sequencer, driver and monitor. Agents can also include components like protocol checkers, TLM model, coverage collectors.

\subsubsection{Sequence Item}
A sequence item is an object modelling the information packet transmitted between two components sometimes referred to as a transaction in the UVM hierarchy. It is written extending sequence\_item class.

\subsubsection{Sequence}
A Sequence is an object that is used to generate a set of sequence items to be sent to the driver.

\subsubsection{Sequencer}
Sequencer takes sequence items from the sequences generated and gives it to the driver. Sequencer and driver use TLM interface functions namely seq\_item\_export and seq\_item\_import to connect with one another.

\subsubsection{Driver}
A Driver is an object which drives the DUT pins based on the sequence items received from the sequencer. It converts the sequences into bitwise values and drives the data onto DUT pins.

\subsubsection{Monitor}
The Monitor takes the DUT bit level values and converts them into sequence items that need to be sent to the other UVM components such as the scoreboard and coverage classes. Monitor uses analysis port for this purpose, and it performs a broadcast operation.

\subsubsection{Scoreboard}
The Scoreboard implements checker functionality. The checker matches the response of DUT with an expected response. It fetches the expected response from a reference module and receives output transactions from the monitor.

A detailed description of all the UVM features can be found in [3].

\section{SPI MASTER CORE}
Serial Peripheral Interface is a communication protocol that facilitates full duplex communication between a master that is usually a microcontroller unit and a slave that is usually a small peripheral device. Communication can happen from master to slave and vice versa. Fig. 2. (S. Simon 2004 [6]) represents the SPI Master Core with three parts [6]. The SPI bus is used to send and receive data between master and slave that could usually be a microcontroller and a small peripheral unit respectively. When compared to other protocols, the SPI protocol has the advantage of relatively high transmission speed, simple to use and uses a small number of signal pins. The protocol divides the devices into master and slave for transmitting and receiving the data. The protocol uses a master device to generate separate clock and data signal lines, along with a chip select line to select the slave device for which the communication has to be established. If there is more than one slave device present, the master device must have multiple slave select interfaces to control the devices.

\begin{figure}[!t]
\centering
\includegraphics[scale=0.5]{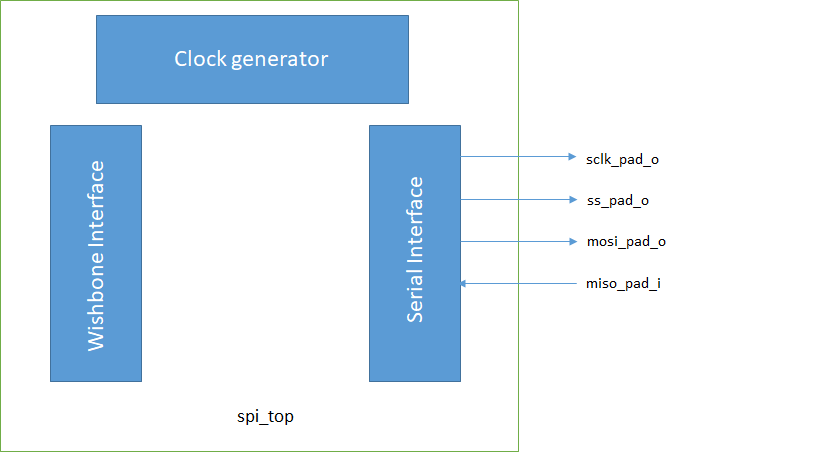}
\caption{SPI Master IP Core}
\end{figure}

There are two data transfer lines. One line responsible for data transfer is Master Out Slave In. The other is Master In Slave Out. Serial Clock (SCLK) is the clock synchronization line, and Slave Select is analogous to chip select. These are the four signals of an SPI bus interface. The configuration of Master Out Slave In (MOSI) line is as an output in a master and as an input in a slave. It facilitates one-way data transmission from master to the slave. The configuration of the Master In Slave Out (MISO) line is as an input in a master and as an output in a slave. It facilitates one-way data transmission from slave to the master. The MISO line remains undriven in a high impedance state when a specific slave is not selected. The Slave Select (SS) is an active low signal that is used as a chip-select line to select a particular slave. The Serial Clock line is used to synchronize data exchange between the MOSI and MISO lines. The required number of bit clock cycles are generated by master depending on the number of bytes transacted between Master and Slave.

The wishbone compliant master core can be seen as organized into three components, namely wishbone interface, clock generator and serial interface [6]. The synthesized RTL view of the clock generator is illustrated in Fig. 3. The serial clock(sclk) is generated by scaling the external system clock(wbclk) with the desired frequency factor as configured by Clock Divider register. The expression of this division is as follows:

\begin{equation}
f_{sclk}=f_{wbclk}/(DIVIDER+1)*2
\end{equation}

\begin{figure}[!t]
\centering
\includegraphics[scale=0.3]{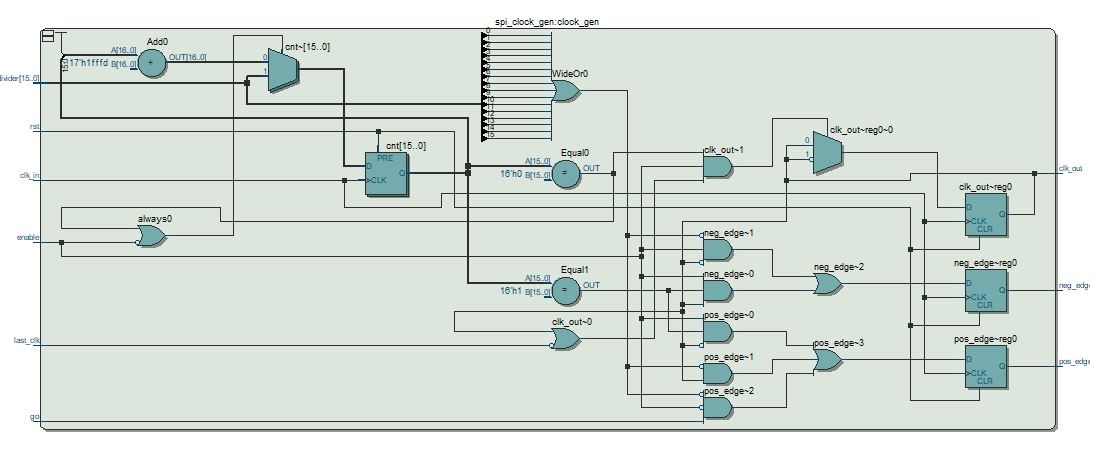}
\caption{Clock generator RTL View}
\end{figure}

The serial data transfer module is responsible for interchanging serial MISO data with parallel data by enforcing appropriate conversion techniques. The top-level module of the SPI Master IP Core has full control over the clock generator and the serial data transfer module.

The SPI master core is our candidate for Design Under Test. We modified an opencores standard SPI Master IP Core design to produce a suitable DUT. The DUT is verified in conjunction with the SPI slave. The approach we took to verify DUT is to send data from both master and slave endpoints. After the transfer is complete, we verify the interchanged data at both ends. Fig. 4 shows the UVM testbench for the DUT. 

The top-level module is responsible for connecting the testbench with the DUT. The environment contains both the agent and the scoreboard. The agent is created using uvm\_agent virtual base class. In the build phase, the agent builds sequencer, driver and monitor components. We enforced randomization on the sequence items. The wishbone bus functional model at the driver side transfers transactional level packets into wishbone specific pin level data.

\section{DEVELOPMENT OF UVM BASED VIP}
The architecture of UVM environment begins with a sequence item. Sequence item is a class object usually extended from uvm\_transaction or uvm\_sequence\_item classes. It consists of all the required data transfers that can be randomized or constrained to the specified boundary by using UVM constructs. Sequences are extended from uvm\_sequence and generate multiple sequence items. The generated sequences are taken to the driver to drive DUT pins. SPI master core driver consists of tasks. The first step in the driver is to get the next sequence item. Secondly, we drive the transfer of data. Thirdly, we write the packet to UVM analysis port, and we are done with the sequence item. The tasks are run simultaneously through a fork...join call. The design of the testbench involves the development of a monitor, which observes the communication of the DUT with the testbench. It observes the pin level transaction at the outputs of the DUT and reports an error if the protocols are violated. The agent connects all these UVM components. The prediction of the DUT's expected output is made in the monitor, and the scoreboard compares the predicted response with the DUT's actual response. The instantiation and connection of agent and scoreboard are done in the env class. 

All UVM classes contain different simulation phases as enumerated here.

\begin{figure}[!t]
\centering
\includegraphics[scale=0.3]{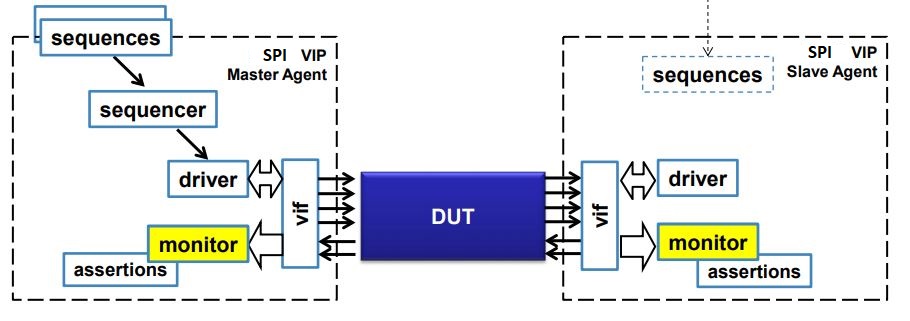}
\caption{UVM testbench}
\end{figure}

\subsection{Build Phase}
The build phase instantiates all the UVM components and is executed at the start of the UVM testbench simulation. 

\subsection{Connect Phase}
The connect phase makes connections among the subcomponents. Testbench connections are made using TLM ports.

\subsection{Elaboration Phase}
Connections are checked in the elaboration phase, address ranges, values and pointers are set up. 

\subsection{Simulation Phase}
Initial runtime configurations are set up in this phase, UVM testbench topology is checked for correctness.

\subsection{Run Phase}
Run phase is the main execution phase where all the simulations are run. This phase starts at time 0.

\subsection{Extract Phase}
This phase extracts data from the DUT and the scoreboard and prepares final statistics.

\subsection{Report Phase}
The Report phase is used to furnish simulation results for the verification engineer's perusal.

\section{RESULTS AND DISCUSSIONS}
Simulations are done and analyzed using QuestaSim. Simulation waveforms are shown in Fig. 5 where we drive the DUT using the Driver component of the VIP.

\begin{figure}[!t]
\centering
\includegraphics[width=3.2in,height=1.7in]{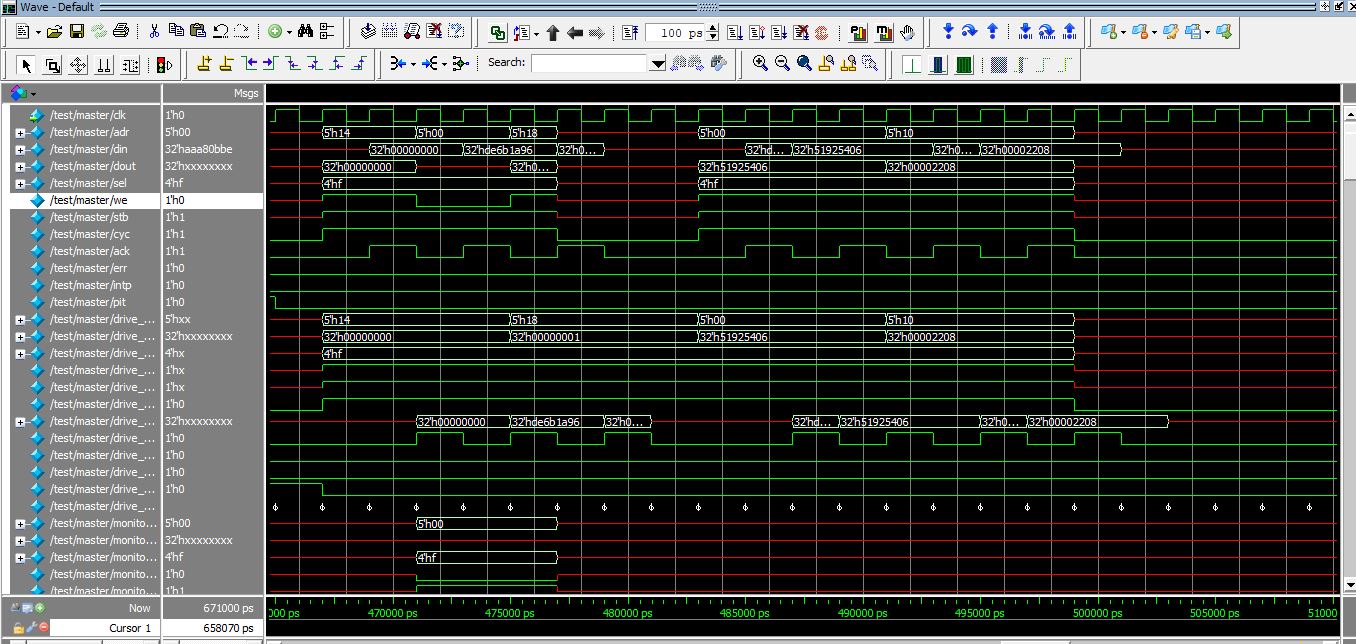}
\caption{Simulation waveform of SPI Master Core}
\end{figure}

While [8] also presents a UVM based verification of SPI Master Core, we have achieved
an improvement over others by adopting constrained random simulation and using an effective set of assertions to capture the designer's intent very well. We propose effective directions for future work in designing highly reliable systems.

Our limitations are as follows: 

\begin{enumerate}

\item At any instance of time, only one master can communicate in SPI.

\item Hot swapping is not supported by SPI, which refers to dynamically adding modules.

\item If at all an error creeps into the protocol, there is no error checking capability like the parity bit built into the protocol.

\item When we operate in a multi-slave model, we would require a separate SS line for each slave, which can get cumbersome when the number of slaves is large.

\item We do not receive an acknowledgement for successful data reception in SPI communication protocol.

\end{enumerate}

The data transfer protocol is illustrated in Fig. 6 (S. Simon 2004 [6]).

\begin{figure}[!t]
\centering
\includegraphics[scale=0.5]{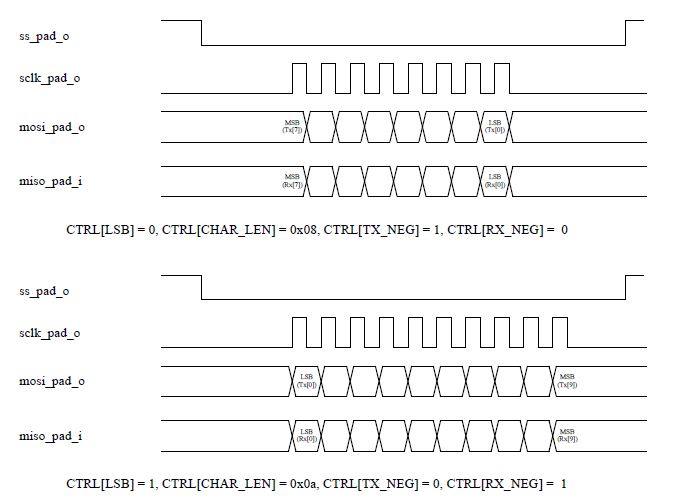}
\caption{Data Transfer Protocol}
\end{figure}

\section{CONCLUSION}
In this paper, we have developed a reusable verification IP for SPI master core that is wishbone compliant. We made use of System Verilog and UVM to propose a reusable testbench that is comprised of Driver, Monitor, SPI slave, scoreboard, agent, environment, coverage analysis and assertions using OOP. Moreover, Constrained Randomization technique is used to achieve wider functional coverage. The results from our simulation-based verification prove the effectiveness of the proposed VIP. This pre-verified SPI Master core IP can be plugged into SOC as it is. This verification component can be reused across the project for verification of other IP. The post verification analysis is done in IMC, a coverage analysis tool that returned a coverage of 92.31\%. 

Simulation-based verification is effective for verifying large systems, but does not give a guaranteed proof of correctness. On the other hand, formal verification gives us a guaranteed proof of correctness by exhaustively covering the state space to unravel corner case bugs but does not scale well, at some point verification gets cumbersome due to state space explosion problem.

In future, consider using formal verification as a complementary step to simulation to improve the confidence of the verified system. The idea is to inject formal analysis into simulation environments. This has been referred by different names, but we prefer to use the name Directed Explicit Model Checking [9]. We start with an A* algorithm and define some heuristics (details of the heuristic is out of the scope of this work) to propagate our search in the direction of a particular failure situation. Then we employ model checking with this particular state as our starting state. As a result of this, counterexamples will be shorter and the state space explored will be smaller, thereby solving our state space explosion problem to a great extent. In [10] also authors propose a Simulation-Guided Formal Analysis approach to bridge the gap between formal techniques and simulation-based methods by leveraging the data obtained from simulations.

\end{document}